\documentclass[apl,twocolumn,amsmath,amssymb,superscriptaddress,floatfix]{revtex4-1}

\usepackage{graphicx}
\usepackage{dcolumn}
\usepackage{bm}
\usepackage[sort&compress]{natbib}
\usepackage{xspace}
\usepackage{amsmath}
\usepackage{epstopdf}
\usepackage{amssymb}
\usepackage{epsfig}

\newcommand{\vecb}[1]{\mathbf{#1}}

\newcommand{\ket}[1]{\left| #1 \right>} 
\newcommand{\bra}[1]{\left< #1 \right|} 

\begin{document}

\title{Moving boundary and photoelastic coupling in GaAs optomechanical resonators}

\author{Krishna C. Balram} \email{krishna.coimbatorebalram@nist.gov}
\affiliation{Center for Nanoscale Science and Technology,
National Institute of Standards and Technology, Gaithersburg, MD
20899, USA}
\affiliation{Maryland NanoCenter, University of Maryland, College
Park, MD}
\author{M. Davan\c co}
\affiliation{Center for Nanoscale Science and Technology,
National Institute of Standards and Technology, Gaithersburg, MD
20899, USA}
\author{Ju Young Lim}
\affiliation{Center for Opto-Electronic Convergence Systems, Korea Institute of Science and Technology, Seoul 136-791, Republic of Korea}
\author{Jin Dong Song}
\affiliation{Center for Opto-Electronic Convergence Systems, Korea Institute of Science and Technology, Seoul 136-791, Republic of Korea}
\author{K. Srinivasan} \email{kartik.srinivasan@nist.gov}
\affiliation{Center for Nanoscale Science and Technology, National
Institute of Standards and Technology, Gaithersburg, MD 20899, USA}

\date{\today}

\begin{abstract}
Chip-based cavity optomechanical systems are being considered for applications in sensing, metrology, and quantum information science.  Critical to their development is an understanding of how the optical and mechanical modes interact, quantified by the coupling rate $g_{0}$.  Here, we develop GaAs optomechanical resonators and investigate the moving dielectric boundary and photoelastic contributions to $g_{0}$. First, we consider coupling between the fundamental radial breathing mechanical mode and a 1550~nm band optical whispering gallery mode in microdisks.  For decreasing disk radius from $R=5$~$\mu$m to $R=1$~$\mu$m, simulations and measurements show that $g_{0}$ changes from being dominated by the moving boundary contribution to having an equal photoelastic contribution.  Next, we design and demonstrate nanobeam optomechanical crystals in which a $2.5$~GHz mechanical breathing mode couples to a 1550~nm optical mode predominantly through the photoelastic effect. We show a significant (30~$\%$) dependence of $g_{0}$ on the device's in-plane orientation, resulting from the difference in GaAs photoelastic coefficients along different crystalline axes, with fabricated devices exhibiting $g_{\text{0}}/2\pi$ as high as 1.1~MHz for orientation along the [110] axis. GaAs nanobeam optomechanical crystals are a promising system which can combine the demonstrated large optomechanical coupling strength with additional functionality, such as piezoelectric actuation and incorporation of optical gain media.
\end{abstract}
\maketitle

\section{Introduction}
\label{sec:Intro}

Mechanical motion and optical fields are coupled by a number of different mechanisms in cavity optomechanical systems~\cite{ref:Aspelmeyer_Kippenberg_Marquardt_Review,ref:Favero_Karrai,ref:Li_Tang_reactive_coupling,ref:Wu_Barclay_dissipative_dispersive_OM}.  Within micro- and nanoscale geometries, perhaps the most commonly considered one is the change in effective optical path length resulting from moving dielectric boundaries, analogous to a movable mirror in a Fabry-Perot cavity. However, the optical path length also depends on the refractive index of the medium filling the cavity, and in solids this can change due to mechanical motion because of the photoelastic effect (electrostriction)~\cite{ref:Rakich_OE_forces_wgs}. This has been observed in stimulated Brillouin scattering in suspended silicon waveguides~\cite{ref:Rakich_Nat_Comm} and cooling and excitation of traveling wave acoustic modes in silica whispering gallery mode resonators~\cite{ref:bahl2011stimulated}.  More recently, it has been considered in silicon optomechanical crystals, where optimized geometries that exclusively rely on the photoelastic effect have been developed~\cite{ref:chan_optimized_OMC}.

In this work, we investigate the moving dielectric boundary and photoelastic contributions to the optomechanical coupling in GaAs devices. GaAs has many desirable properties for cavity optomechanics: relatively large photoelastic coefficients~\cite{ref:Rakich_OE_forces_wgs} which can produce devices with high optomechanical coupling~\cite{ref:chan_optimized_OMC}; piezoelectric properties~\cite{ref:Masmanidis_Roukes_piezoelectric_Science} which can be exploited for driving or readout of mechanical motion; and potential integration with InAs/GaAs quantum dots that offer non-classical light emission and a strong resonant nonlinearity~\cite{ref:Michler_book_2009} that can be used to probe and control mechanical motion~\cite{ref:Wilson-Rae,ref:yeo2013strain}.

\begin{figure*}
\begin{center}
\begin{minipage}[c]{0.66\linewidth}
\includegraphics[width=\linewidth]{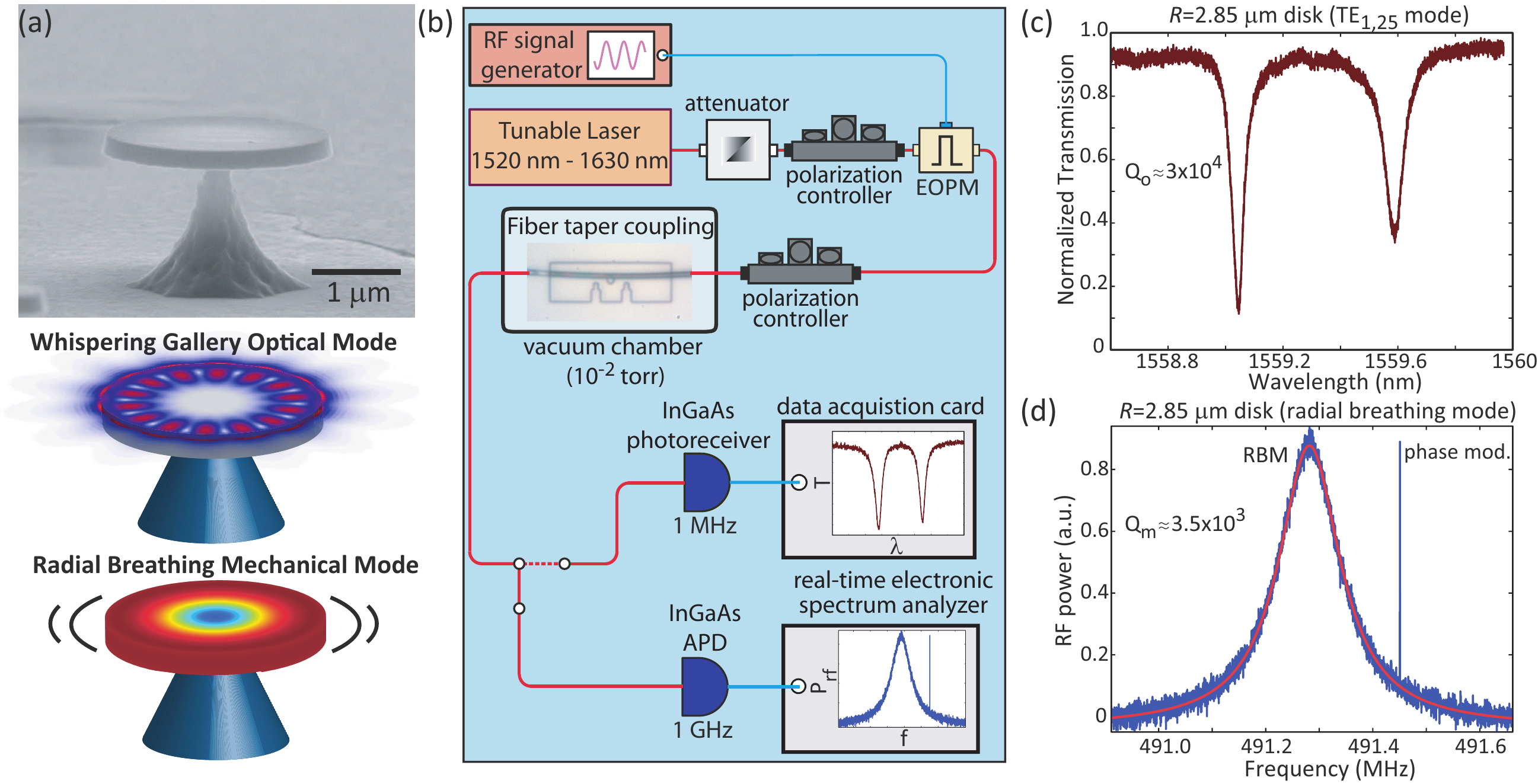}
\end{minipage}\hfill
\begin{minipage}[c]{0.32\linewidth}
\caption{GaAs microdisk optomechanical resonators. (a) Scanning electron microscope image of a fabricated device and finite-element-method
simulations of the optical (TE$_{1,7}$) and mechanical mode (1.4 GHz radial breathing mode) in a $R$=1~$\mu$m microdisk. (b) Experimental setup for measuring the optomechanical coupling. EOPM = electro-optic phase modulator; APD = avalanche photodiode. (c) Optical transmission spectrum for the TE$_{1,25}$ mode in a $R=2.85~\mu$m diameter device. (d) Thermal noise spectrum for the $\approx$~490~MHz radial breathing mode, shown together with Lorentzian fit (red) and the phase modulator calibration peak.}\label{fig:Figure1}
\end{minipage}
\end{center}
\vspace{-0.2in}
\end{figure*}

We first present a combined theoretical and experimental analysis of GaAs microdisks of varying radius, where in large radius devices the moving boundary effect dominates, while in small radius devices the photoelastic effect is the leading contribution. While these trends have recently been theoretically predicted~\cite{ref:Baker2014photoelastic}, here we experimentally demonstrate the importance of considering both these effects in the overall optomechanical coupling rate $g_{0}$. We then present two designs of GaAs nanobeam optomechanical crystals that rely predominantly on the photoelastic effect. We show a significant (30~$\%$) dependence of $g_{0}$ on the in-plane device angle, in contrast with similar Si devices~\cite{ref:chan_optimized_OMC}, for which the dependence is much weaker (3~$\%$).  This dependence originates from the much larger magnitude (and opposite sign) of the photoelastic coefficient $p_{12}$ in GaAs.  We experimentally demonstrate this effect by measuring $g_{0}$ in devices fabricated with differing in-plane angles, and measure $g_{0}/2\pi$ as high as 1.1~MHz for devices oriented along the [110] axis of GaAs.  Mechanical modes at 2.5~GHz with a quality factor $Q_{\text{m}}\approx$~2000 at room temperature and atmosphere are observed, as is self-oscillation of the mechanical modes through radiation-pressure driven dynamical back-action~\cite{ref:Kippenberg_Vahala_OE}.

\section{Microdisks}
\label{sec:udisks}

Figure~\ref{fig:Figure1}(a) shows a scanning electron microscope image of a microdisk cavity fabricated in a 220~nm thick GaAs layer using typical lithography and dry etching processes (see Supplementary Material).  Finite-element-method simulations are used to calculate the whispering gallery optical modes and radial breathing mechanical modes of such devices (Fig.~\ref{fig:Figure1}(a)), with disk radius $R$ varying between 1~$\mu$m and 5~$\mu$m.  For each value of $R$, we calculate optical modes of transverse electric (TE) polarization (dominant electric field components are in the plane of the disk), and determine the azimuthal mode number $m$ that places a first order radial mode in the 1550~nm band.  We focus on the TE$_{(1,m)}$ mode because of its comparatively high radiation-limited optical quality factor ($Q_{o}$) for small disks ($R\gtrsim1$~$\mu$m).  Similarly, we focus on the fundamental radial breathing mechanical mode as it is expected to have a higher mechanical quality factor ($Q_{m}$) than higher-order modes for a given supporting pedestal size.

The optomechanical coupling rate $g_{0}$, defined as the optical mode frequency shift due to the mechanical mode's zero-point motion~\cite{ref:Gorodetsky_Kippenberg_OM}, has moving boundary ($g_{0,\text{MB}}$) and photoelastic ($g_{0,\text{PE}}$) contributions obtained from the calculated modes as in Ref.~\cite{ref:chan_optimized_OMC}:

\begin{align}
g_{0,MB}=&-\frac{\omega_{0}}{2}\frac{\oint_{A}dA(\vecb{Q}\cdot\vecb{\hat{n}})(\Delta\epsilon|\vecb{E}_{||}|^{2}-\Delta\epsilon^{-1}|\vecb{D}_{\bot}|^{2})}{\int dV\epsilon|\vecb{E}|^{2}}\\
g_{0,PE}=&-\frac{\omega_{0}\epsilon_{0}n^{4}}{2}\frac{\int dV\sum(|\vecb{E}|^{2}(p_{11}S_{xx}+p_{12}(S_{yy}+S_{zz}))}{\int dV\epsilon|\vecb{E}|^{2}} \nonumber\\
  &-\frac{\omega_{0}\epsilon_{0}n^{4}}{2}\frac{\int dV\sum(|\vecb{E}|^{2}4Re(E_{x}^{*}E_{y})p_{44}S_{xy})}{\int dV\epsilon|\vecb{E}|^{2}}
\label{eq:g0_eqns}
\end{align}

\noindent where $\sum$ is a summation according to the Einstein notation $x\rightarrow y\rightarrow z\rightarrow x$. The $p_{ii}$ are the photoelastic coefficients of GaAs ($p_{11}=-0.165$, $p_{12}=-0.14$, $p_{44}=-0.072$), $S_{ii}$ is the strain, and $\vecb{Q}$ is normalized mechanical displacement.  Qualitatively, the photoelastic contribution is sensitive to mechanical motion throughout the device, whereas the moving boundary contribution is sensitive to the motion of surfaces (particularly the disk sidewall for the radial breathing mode).

\begin{figure}[t]
\centerline{\includegraphics[width=\linewidth]{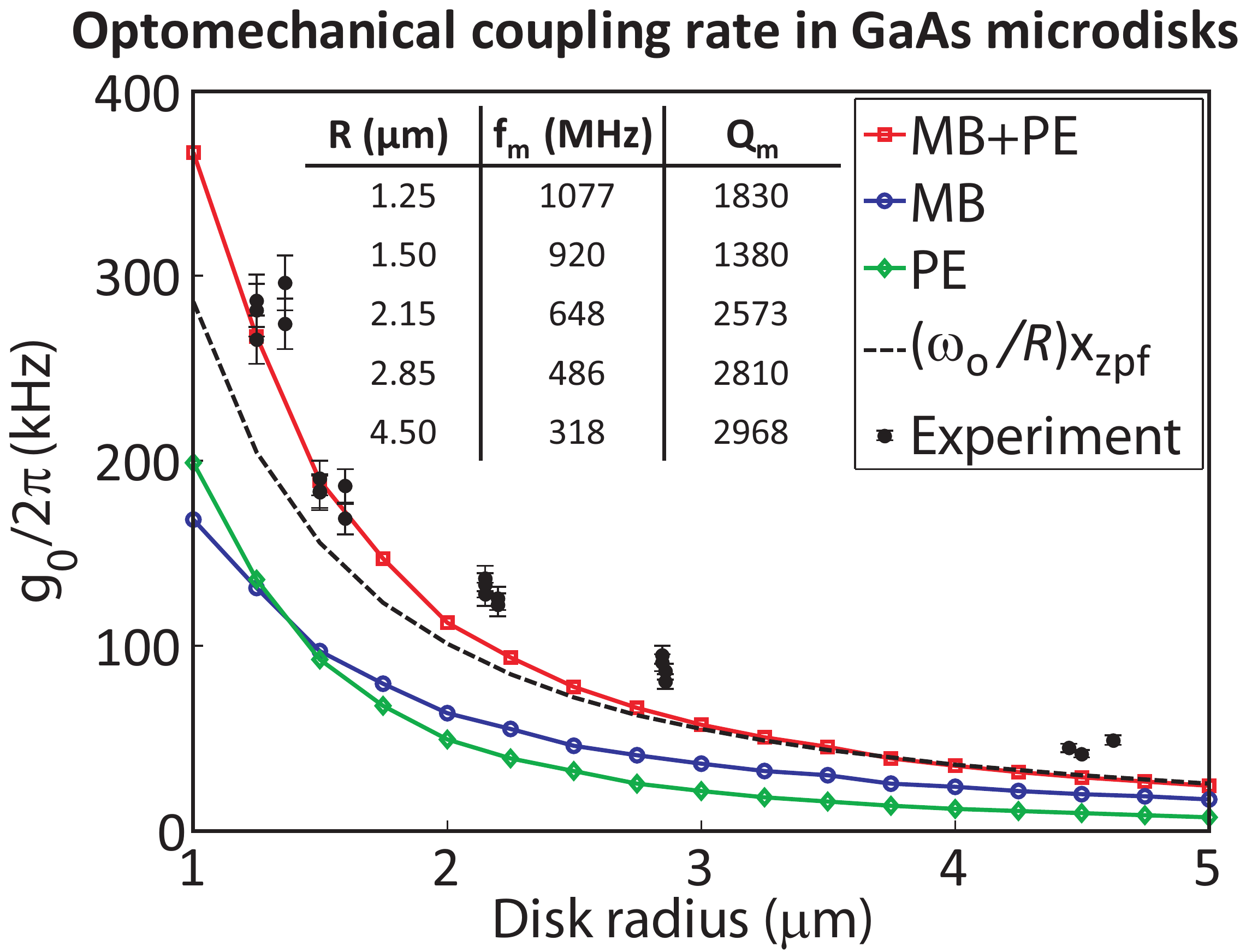}}
 \caption{Optomechanical coupling rate $g_{\text{0}}$ as a function of radius $R$, for coupling between the TE$_{(1,m)}$ optical modes and fundamental radial breathing mechanical modes.  Red, blue, and green curves are the calculated total coupling rate (MB+PE), moving boundary (MB) contribution, and photoelastic (PE) contribution, respectively.  Dashed black line is a rough estimate $g_{\text{0}}=(\omega_{o}/R)x_{\text{zpf}}$, where $\omega_{\text{o}}$ is the optical frequency and $x_{\text{zpf}}$ is the zero-point motion amplitude. Black circles are experimental values, where the error bars are dominated by uncertainty in the modulator $V_{\pi}$ and are one standard deviation values.  Inset table gives the measured mechanical frequency and $Q_{\text{m}}$.}
\label{fig:Figure2}
\end{figure}

Figure~\ref{fig:Figure2} shows the calculated contribution to $g_{0}$ due to the moving boundary (blue) and photoelastic effects (green) as a function of $R$. For $R\gtrsim2$~$\mu$m, the moving boundary effect dominates whereas for $R\lesssim2$~$\mu$m, the photoelastic effect is comparable or even slightly larger.  This is consistent with recent simulation results for similar GaAs microdisks~\cite{ref:Baker2014photoelastic}. To verify this scaling behavior experimentally (Fig.~\ref{fig:Figure1}(b)), we measure fabricated GaAs microdisks of varying radius, following an approach similar to Ref.~\cite{ref:Gorodetsky_Kippenberg_OM}, where a calibration signal of known modulation index $\beta_{pm}$ from a phase modulator driven close to the mechanical resonance frequency is used to determine the magnitude of $g_{0}$ (see Supplementary Material). Figure~\ref{fig:Figure1}(c)-(d) shows representative optical and mechanical modes for a device, where the mechanical mode spectrum also includes the phase modulator calibration tone.  Compared to microdisks fabricated previously using an essentially identical process~\cite{ref:Srinivasan9}, $Q_{o}$ in these devices ($<5{\times}10^4$) is an order of magnitude lower, likely as a result of $100$~nm length scale roughness present on the underside of the GaAs layer (see Supplementary Material).  Such lower $Q_{o}$ values do not influence the estimate of $g_{0}$, but do prevent operation in the sideband-resolved regime needed for a number of applications.

The optomechanical coupling rate $g_{0}$ can be estimated from a mechanical mode spectrum as (see Supplementary Material):

\begin{equation}
g_{0}^{2}=\frac{\hbar\Omega_{m}}{2k_{B}T}\Omega_{m}^{2}\beta_{pm}^{2}\frac{S_{cav}(\Omega_{m})}{S_{pm}(\Omega_{mod})}
\end{equation}

\noindent where $S_{cav}(\Omega_{m})$ is the power in the mechanical mode, $S_{pm}(\Omega_{mod})$ is the power in the phase modulator signal, and $\beta_{pm}$ is the modulation index. $\beta_{pm}=\pi\frac{V_{sig}}{V_{\pi}}$, where $V_{sig}$ is the applied voltage and $V_{\pi}$ (the voltage required to produce a $\pi$ phase shift) is determined through a separate calibration (see Supplementary Material).

The experimentally measured $g_{0}$ values for disks of varying $R$ are plotted in Fig.~\ref{fig:Figure2} (black circles), where the uncertainty in each measurement is dominated by the uncertainty in the phase modulator $V_{\pi}$. The data shows good agreement with the red curve, which plots the sum of the moving boundary and the photoelastic contributions to $g_{0}$. Especially for small disk radii, the data shows significant deviation from the moving boundary contribution alone, which might help explain some discrepancies observed in previous measurements of GaAs disk optomechanical resonators, where only moving boundary effects were considered in simulation comparisons~\cite{ref:Ding_Favero_GaAs_disk_optomechanics}. We note that for nominally identical disks, a spread in $g_{0}$ of $\approx10~\%$ is observed.  We attribute this to the specifics of the fiber taper coupling for each device, which, we have observed, can perturb the optical and mechanical modes and the resultant optomechanical coupling (see Supplementary Material).

\section{Nanobeam optomechanical crystals}
\label{sec:nanobeams}

The measurements of $g_{0}$ in GaAs microdisks demonstrated that the contribution due to the photoelastic effect ($g_{0,PE}$) can become comparable to, and even exceed that due to the moving boundary ($g_{0,MB}$) effect, as the disk radius $R$ becomes comparable to the wavelength. Intuitively, as $R$ is reduced, the volume of both the optical and mechanical modes decrease and $g_{0}$ increases due to increased spatial overlap. Given that bending loss starts to dominate $Q_{o}$ for $R<0.7$~$\mu$m, one can estimate that $g_{0}/2\pi\lesssim450$~kHz based on the data shown in Fig.~\ref{fig:Figure2}. For higher $g_{0}$, one needs to consider geometries that support more tightly confined optical and mechanical modes.

\begin{figure*}
\begin{center}
\includegraphics[width=\linewidth]{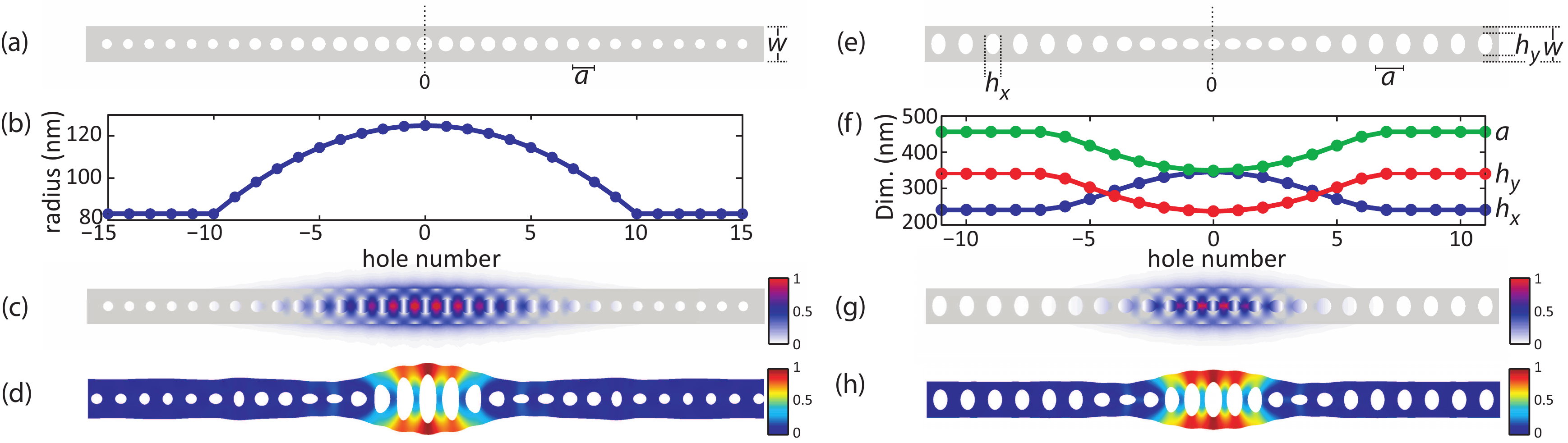}
\caption{GaAs nanobeam optomechanical crystal designs based on (a)-(d) a circular hole geometry and (e)-(h) an elliptical hole geometry.  For each, we plot the design ((a) and (e)), variation in design parameters as a function of hole number ((b) and (f)), normalized electric field amplitude ((c) and (g)), and normalized mechanical displacement ((d) and (h)). }\label{fig:Figure3}
\end{center}
\end{figure*}

Optomechanical crystals~\cite{ref:eichenfield2}, structures that spatially co-localize optical modes within a photonic bandgap and mechanical modes within a phononic bandgap, have been demonstrated in a number of materials~\cite{ref:chan_optimized_OMC,ref:fan_hong_tang,ref:Bochmann_Cleland_AlN_uwave_optomechanics,ref:Davanco_nanobeam_OMC,ref:kipfstuhl2014modeling,ref:Phoxonic_crystals_Rolland} and in both one- and two-dimensional geometries.  Here, we focus on one-dimensional (nanobeam) geometries, where the nanobeams are patterned with a series of holes whose dimensions are graded quadratically from the center (cavity section) to the edge (mirror section). The general design principle for the optical cavity~\cite{ref:Quan1} relies on choosing the center hole dimension to support a guided mode and then quadratically tapering the hole dimension down to the mirror section where the mode lies in the forbidden band and hence is reflected. The cavity is constructed by putting two such tapers back-to-back. The quadratic taper ensures that the electric field amplitude of the mode has a Gaussian profile and retains high $Q_{o}$.  Similar design concepts have been used in the development of GaAs nanobeam photonic crystal cavities for applications in lasing~\cite{ref:Gong_Vuckovic_OE_nanobeam} and cavity quantum electrodynamics~\cite{ref:Ohta_cQED_nanobeam}.  Here, we demonstrate these GaAs cavities in the context of cavity optomechanics, where we note that the tailoring of the hole dimensions also enables localization of mechanical modes~\cite{ref:eichenfield2}.

Figure~\ref{fig:Figure3} shows two different nanobeam optomechanical crystal cavity designs in 220~nm thick GaAs, using circular holes (Fig.~\ref{fig:Figure3}(a)-(d)) as described above and elliptical holes (Fig.~\ref{fig:Figure3}(e)-(h)) in an approach similar to Ref.~\cite{ref:chan_optimized_OMC}.  For each design, we show a schematic of the geometry, the variation in device parameters as a function of hole number, and the optical and mechanical modes. We designed the devices to have a nominal operating wavelength of 1550 nm and ran a parameter sweep to find designs with the highest $g_{0}Q_{o}$, where both moving boundary and photoelastic contributions to $g_{0}$ are calculated as described earlier for microdisks.

The circular hole design is one realization of the more general elliptical hole design, and is thus comparatively simple in terms of the number of design parameters.  In particular, we fix the lattice constant $a$ and adjust only the hole radius $r$ at the different lattice sites.  More specifically, we vary the hole radius in the center of the cavity and the mirror sections, as well as the steepness of the quadratic grading profile (i.e., the number of holes over which the radius is tapered).  In comparison, the elliptical hole designs have quadratic grades for the lattice constant ($a$) and lengths of the principal axes of the ellipse ($h_{x}$ and $h_{y}$).  Thus, for each dimension $a$, $h_{x}$, and $h_{y}$, we vary the value in center of the cavity and mirror sections, as well as the steepness of the quadratic grade.

While for both the circular and elliptical hole designs, we find parameters for which $Q_{o}>10^6$, $g_{0}$ is higher for the elliptical hole designs.  In particular, the optimized elliptical hole design has $g_{0,PE}/2\pi=860$~kHz and $g_{0,MB}/2\pi=-94$~kHz for coupling between the $\lambda\approx1535$~nm optical mode and $\Omega_{m}/2\pi\approx$~2.14~GHz mechanical mode, compared to the optimized circular hole design with $g_{0,PE}/2\pi=563$~kHz and $g_{0,MB}/2\pi=-43$~kHz for coupling between the $\lambda\approx1545$~nm optical mode and $\Omega_{m}/2\pi\approx2.31$~GHz mechanical mode. The elliptical design ensures a higher $g_{0}$ by having a higher GaAs volume fraction in the center of the beam. We note that while in microdisks, the moving boundary and photoelastic contributions to $g_{0}$ are comparable in magnitude and of the same sign, for the optimized nanobeams, $g_{0}$ is dominated by the photoelastic effect and the moving boundary contribution is opposite in sign and thus reduces the net optomechanical coupling rate.  This behavior is consistent with Refs.~\cite{ref:chan_optimized_OMC} and ~\cite{ref:kipfstuhl2014modeling}, where an optimized nanobeam geometry dominated by the photoelastic effect was developed for Si and diamond, respectively.

\begin{figure}[h]
\centerline{\includegraphics[width=\linewidth]{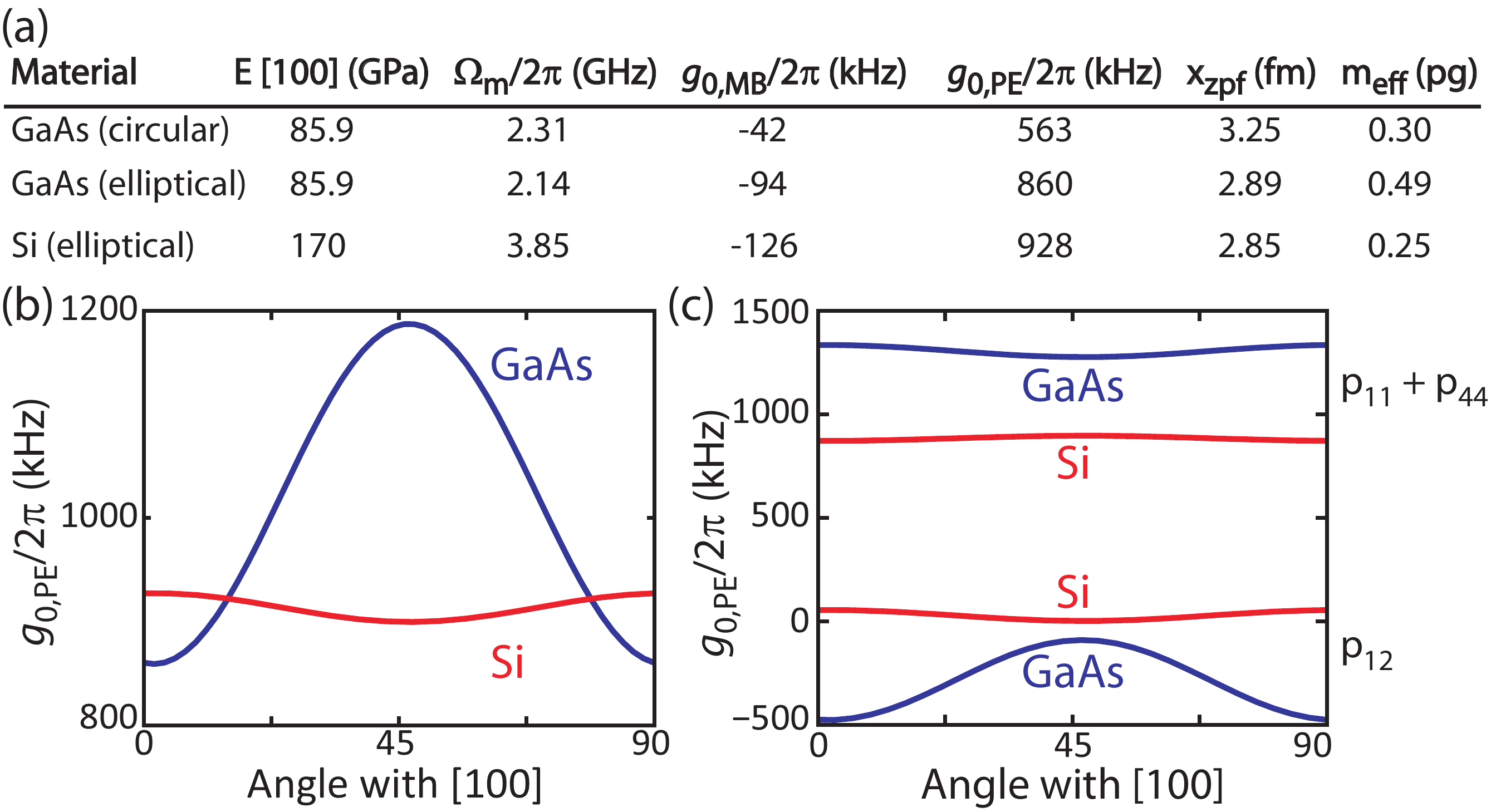}}
 \caption{(a) Parameters for GaAs and Si nanobeam optomechanical crystal designs, including Young's modulus along [100], mechanical mode frequency, optical wavelength, $g_{0,MB}$ and $g_{0,PE}$ (nanobeam long axis oriented along the [100] direction), zero-point motional amplitude, and effective motional mass.  (b) Dependence of $g_{0,PE}$ for the GaAs and Si elliptical designs on in-plane rotational angle. (c) Breakdown of $g_{0,PE}$ into $p_{11}+p_{44}$ (top) and $p_{12}$ (bottom) terms.}
\label{fig:Figure4}
\end{figure}

Given that the photoelastic effect is represented by a tensor, one would expect the contribution to $g_{0}$ to depend on the in-plane orientation of the nanobeam. Moreover, the fact that $p_{12}$ has a much larger magnitude in GaAs than in Si ($p_{12,GaAs}=-0.14$, $p_{12,Si}=0.017$) suggests that the dependence of $g_{0}$ on in-plane orientation will be much more significant in GaAs.  Figure~\ref{fig:Figure4}(b) shows $g_{0,PE}$ for the elliptical hole nanobeam shown in Fig.~\ref{fig:Figure3}(e)-(h) as a function of the angle the long axis of the nanobeam makes with the [100] direction. The coupling rate is calculated using the rotated photoelastic tensor (see Supplementary Material). We can see that the effect is quite significant with a variation of almost 35~\% and a peak value $g_{0,PE}/2\pi\approx$~1.2~MHz at 45~$^{\circ}$ (device orientation along [110]). We also plot the orientation dependence of $g_{0,PE}$ for a silicon nanobeam optomechanical crystal similar to that of Ref.~\cite{ref:chan_optimized_OMC}.  We see that again, there is a dependence of $g_{0,PE}$ on in-plane orientation, although in this case the optomechanical coupling rate is minimized at 45~$^{\circ}$ and the variation between 0~$^{\circ}$ and 45~$^{\circ}$ is less than 5~\%. To understand this, we plot in Fig.~\ref{fig:Figure4}(c) the contributions due to the $(p_{11}+p_{44})$ and $p_{12}$ terms of eqn.~\ref{eq:g0_eqns} as a function of in-plane angle. The in-plane anisotropy of $g_{0,PE}$ is seen to arise primarily from the contribution due to the $p_{12}$ term, and the aforementioned difference in the $p_{12}$ values for GaAs and Si helps us understand why this rotational dependence is so weak in Si. We also point out that the in-plane anisotropy of the elastic coefficients of GaAs leads to an orientation-dependent Young's modulus~\cite{ref:brantley_E_fn_angle,ref:Hopcroft_Youngs_Modulus_Si} and a resultant shift in the mechanical mode frequency (see Supplementary Material), but the moving boundary contribution $g_{0,MB}$ is relatively insensitive to in-plane orientation under the assumption of GaAs being an isotropic elastic material (as can be seen from the form of eqn.~1).  The total optomechanical coupling rate for the GaAs elliptical hole design, including both photoelastic and moving boundary contributions, is therefore as high as $g_{0}/2\pi\approx$~1.1~MHz.

\begin{figure*}
\begin{center}
\includegraphics[width=0.625\linewidth]{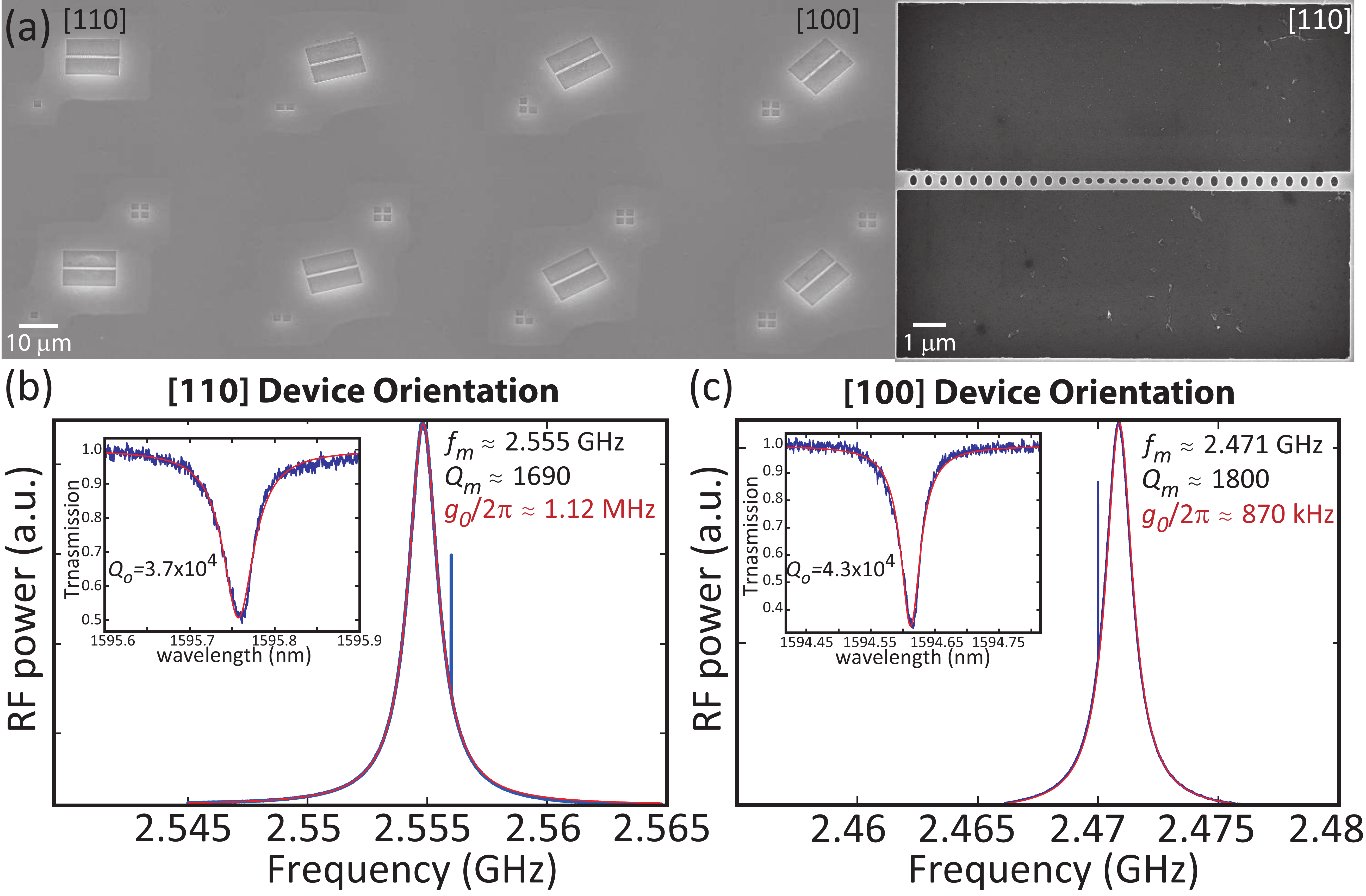}
\caption{GaAs nanobeam optomechanical crystal measurements. (a) Scanning electron microscope image of an array of fabricated devices, where the orientation of the nanobeam long axis is varied between [110] and [100]. The right image is zoomed-in on a single nanobeam cavity aligned along the [110] axis. (b) Thermal noise spectrum (blue curve) and Lorentzian fit (red curve) for a nanobeam breathing mode when the device is aligned along the [110] axis.  The phase modulator calibration approach is used to extract the optomechanical coupling rate $g_{0}/2\pi=1.12$~MHz~$\pm$~0.06~MHz. The inset shows the transmission spectrum (blue) and fit (red) for the nanobeam optical mode. (c) Thermal noise spectrum (blue curve) and Lorentzian fit (red curve) for a nanobeam breathing mode when the device is aligned along the [100] axis.  The phase modulator calibration approach is used to extract the optomechanical coupling rate $g_{0}/2\pi=870$~kHz~$\pm$~45~kHz. The inset shows the transmission spectrum (blue) and fit (red) for the nanobeam optical mode.  The uncertainty values in $g_{0}$ are dominated by uncertainty in the modulator $V_{\pi}$ and are one standard deviation values.}\label{fig:Figure5}
\end{center}
\end{figure*}

We study this orientation-dependent optomechanical coupling by fabricating a series of GaAs nanobeam devices according to the elliptical hole geometry described above, while varying the orientation of the long axis of the nanobeam between [110] and [100], as shown in Fig.~\ref{fig:Figure5}(a).  Devices are tested in the same measurement setup used to test the microdisk samples (Fig.~\ref{fig:Figure1}(b)). In this case, the sample is kept at atmospheric pressure, as gas damping is expected to have limited influence on the mechanical modes due to their high frequencies~\cite{ref:Ekinci_fluid_damping_mechanical_resonators}, an effect that has also been observed recently in Si$_3$N$_4$ nanobeam optomechanical crystals~\cite{ref:Davanco_nanobeam_OMC}.  Optical modes in the 1550~nm band are observed with typical quality factors $Q_{o}\approx 4.0\times10^4$ (Fig.~\ref{fig:Figure5}(b)-(c)). The corresponding mechanical mode frequencies vary between $\approx$~2.555~GHz (alignment along [110]) and $\approx$~2.471~GHz (alignment along [100]), with the $\approx$~85~MHz shift resulting primarily from the anisotropy in the Young's modulus of GaAs (anisotropy in the Poisson's ratio and shear modulus are also included in simulations; see Supplementary Material).  Mechanical quality factors are typically around $Q_{m}\approx 2\times$10$^3$.

\begin{figure}[h!]
\centerline{\includegraphics[width=\linewidth]{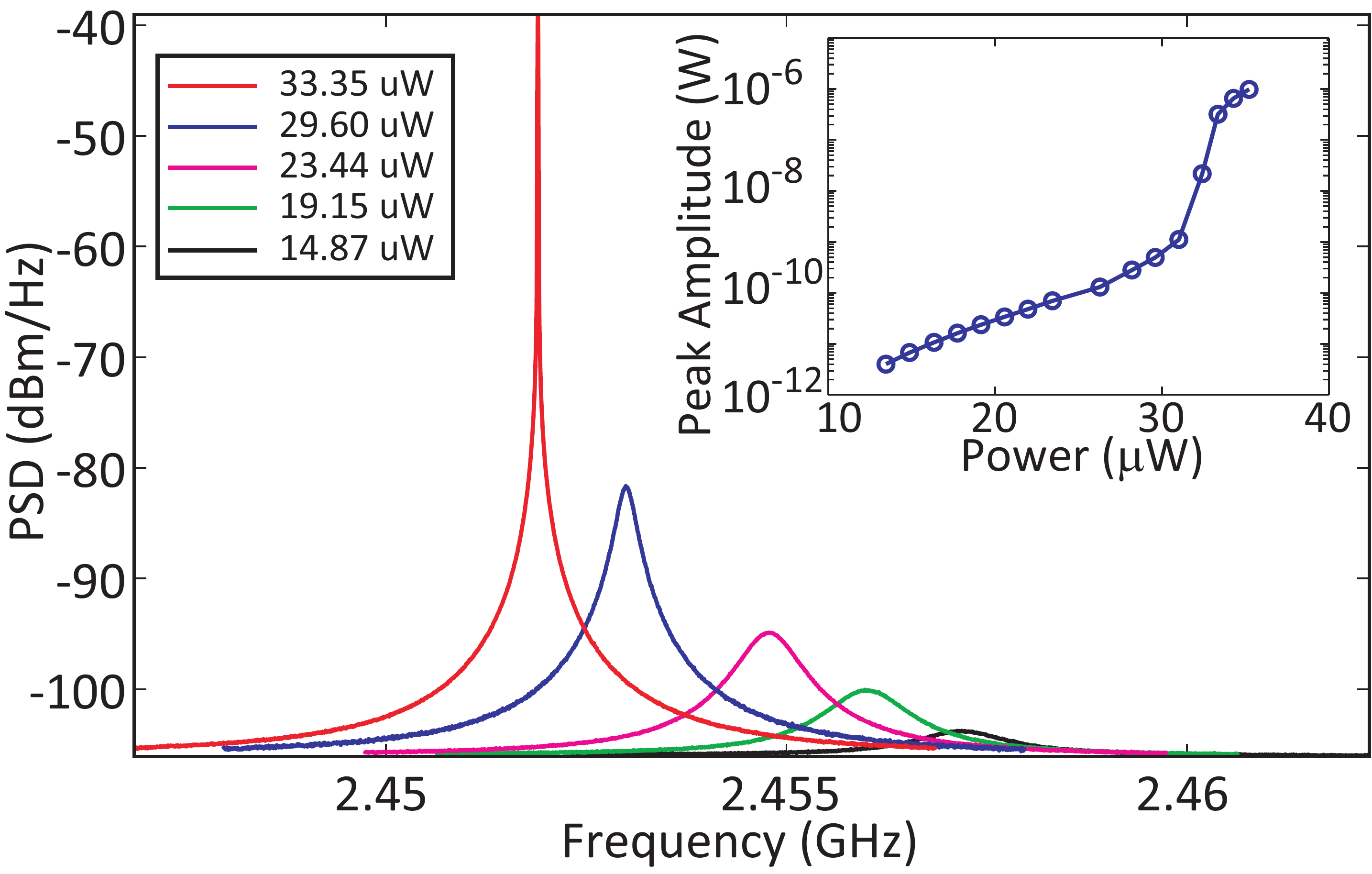}}
 \caption{(a) Mechanical mode spectra as a function of increasing optical power injected into a GaAs nanobeam optomechanical crystal aligned along the [100] axis, showing a pronounced linewidth narrowing and peak height increase. (b) Mechanical mode peak amplitude as a function of injected optical power, showing a clear threshold behavior indicative of the system being driven into regenerative mechanical oscillation.  The uncertainty in the peak amplitude is less than the data point size.}
\label{fig:Figure6}
\end{figure}

Using the phase modulator calibration approach described in Section~\ref{sec:udisks}, we extract the optomechanical coupling rate for the different devices, with $g_{0}/2\pi=870$~kHz~$\pm$~45~kHz for the device aligned along [100] (Fig.~\ref{fig:Figure5}(c)) and $g_{0}/2\pi=1.12$~MHz~$\pm~$0.06~MHz for the device aligned along [110] (Fig.~\ref{fig:Figure5}(b)). These values correspond reasonably well with simulations, which when taking into account both the photoelastic (Fig.~\ref{fig:Figure4}(b)) and moving boundary (Fig.~\ref{fig:Figure4}(a)) contributions, predict $g_{0}/2\pi=770$~kHz and $g_{0}/2\pi=1.09$~MHz, respectively.  Measurement of devices fabricated at an intermediate angle of 15~$^{\circ}$ with respect to [100] (not shown) yield $g_{0}/2\pi=920$~kHz~$\pm$~50~kHz, respectively, which also matches reasonably well with the simulation result of $g_{0}/2\pi=850$~kHz.  We note that while the uncertainty values we have quoted for the measured $g_{0}$ are the one standard deviation value due to the uncertainty in the phase modulator $V_{\pi}$, another source of uncertainty is in the precise angle with which the cavity was fabricated relative to the GaAs crystal planes (i.e., the alignment of the GaAs chip within the electron-beam lithography system).  In particular, for intermediate device orientation angles between [100] and [110], Fig.~\ref{fig:Figure4}(b) predicts a variation in $g_{0}/2\pi$ of $\approx$~50~kHz for a $5~^{\circ}$ offset in orientation.  We also note that the $g_{0}$ simulations presented in Fig.~\ref{fig:Figure4} assumed that GaAs is an isotropic elastic material, which led to the aforementioned invariance of $g_{0,MB}$ on in-plane orientation. In the Supplementary Material, we present simulations that treat GaAs as an orthotropic elastic material, and calculate $g_{0,MB}$ and $g_{0,PE}$ for orientation along [100] and [110] by rotating the structure within the simulation, rather than using the rotated photoelastic tensor.  Since $g_{0,MB}$ now has some orientation dependence, we find the calculated values for $g_{0}$ to be a bit closer to the experimental results.

Finally, by injecting increasing levels of optical power into the devices while keeping the laser frequency blue-detuned and on the shoulder of the optical cavity mode, we can drive the system into regenerative mechanical oscillation~\cite{ref:Kippenberg_Vahala_OE}.  Figure~\ref{fig:Figure6}(a) shows a series of mechanical mode spectra for a device oriented along [100], as a function of increasing optical power, from which a clear linewidth narrowing and peak amplitude increase are observed.  The peak amplitude is plotted as a function of input optical power in Fig.~\ref{fig:Figure6} (b).  Here, a characteristic threshold behavior at an input power $<35$~$\mu$W is seen, indicating that the system is indeed self-oscillating.

In summary, we have compared the moving dielectric boundary and photoelastic contributions to the optomechanical coupling rate in GaAs optomechanical resonators.  Simulations and experiments in microdisk cavities correspond closely and show that these two effects have near equal magnitude for devices with a radius near 1~$\mu$m.  Simulations and experiments on nanobeam optomechanical crystals optimized for photoelastic coupling (one order of magnitude larger than the moving boundary effect) show a significant dependence on the in-plane orientation of the nanobeam, with overall coupling rates $g_{0}/2\pi=1.1$~MHz achieved for coupling between 1550~nm optical modes and 2.5~GHz mechanical modes.

\section{Funding Information}
This work was partially supported by the DARPA MESO program. K.C.B. acknowledges support under the Cooperative Research Agreement between the University of Maryland and NIST-CNST Award 70NANB10H193. J.Y.L. and J.D.S. acknowledge support from KIST institutional programs, including the flag-ship, future convergence pioneer, and GRL programs.

\section{Acknowledgments}

We acknowledge J. Kim and S. Krishna from the Center of High Technology Materials at the University of New Mexico for providing epitaxial GaAs material used in fabricating microdisk devices.  J.Y.L. is now with the Laser Technology Research Center, Korea Photonics Technology Institute.

\onecolumngrid \bigskip
\appendix
\setcounter{figure}{0}
\setcounter{equation}{0}
\makeatletter
\renewcommand{\theequation}{S\@arabic\c@equation}
\begin{center} {{\bf \large SUPPORTING
INFORMATION}}\end{center}

\section*{Fabrication Procedure}

The epitaxial material used in the microdisk measurements consists of a 220~nm thick GaAs layer on a 1.5~$\mu$m thick Al$_{0.6}$Ga$_{0.4}$As sacrificial layer. The samples were spin-coated with positive tone electron beam resist and baked at $180\,^{\circ}{\rm C}$ for 2 minutes. The microdisk patterns were exposed in a 100~keV direct write electron beam lithography system with a beam current of 200 pA and nominal dose of 250~$\mu$C/cm$^2$. After exposure, the electron beam resist was developed using hexyl acetate (65 sec). For improved sidewall roughness, the resist was reflowed at $140\,^{\circ}{\rm C}$ for 1 min. The microdisk patterns were then transferred to the underlying GaAs layer using an inductively coupled plasma reactive ion etcher with an Ar/Cl$_2$ chemistry. The electron beam resist was stripped using trichloroethylene, and the microdisks were undercut with a timed wet etch (depending on the disk radius) using (NH$_4$)$_2$S and dilute HF~\cite{ref:perahia2009_thesis}.

The epitaxial material used in the nanobeam optomechanical crystal measurements consists of a 220~nm thick GaAs layer on a 1.5~$\mu$m thick Al$_{0.7}$Ga$_{0.3}$As sacrificial layer. Device fabrication follows the same general procedure as with the microdisks, with two changes: (1) a slightly higher Cl$_2$ percentage is used in the Ar/Cl$_2$ etch to achieve near-vertical sidewalls in the nanobeam holes, and (2) the devices are undercut with 49~$\%$ HF.

\section*{Gallium Arsenide underside roughness}

\begin{figure}[h]
\centerline{\includegraphics[width=0.5\linewidth]{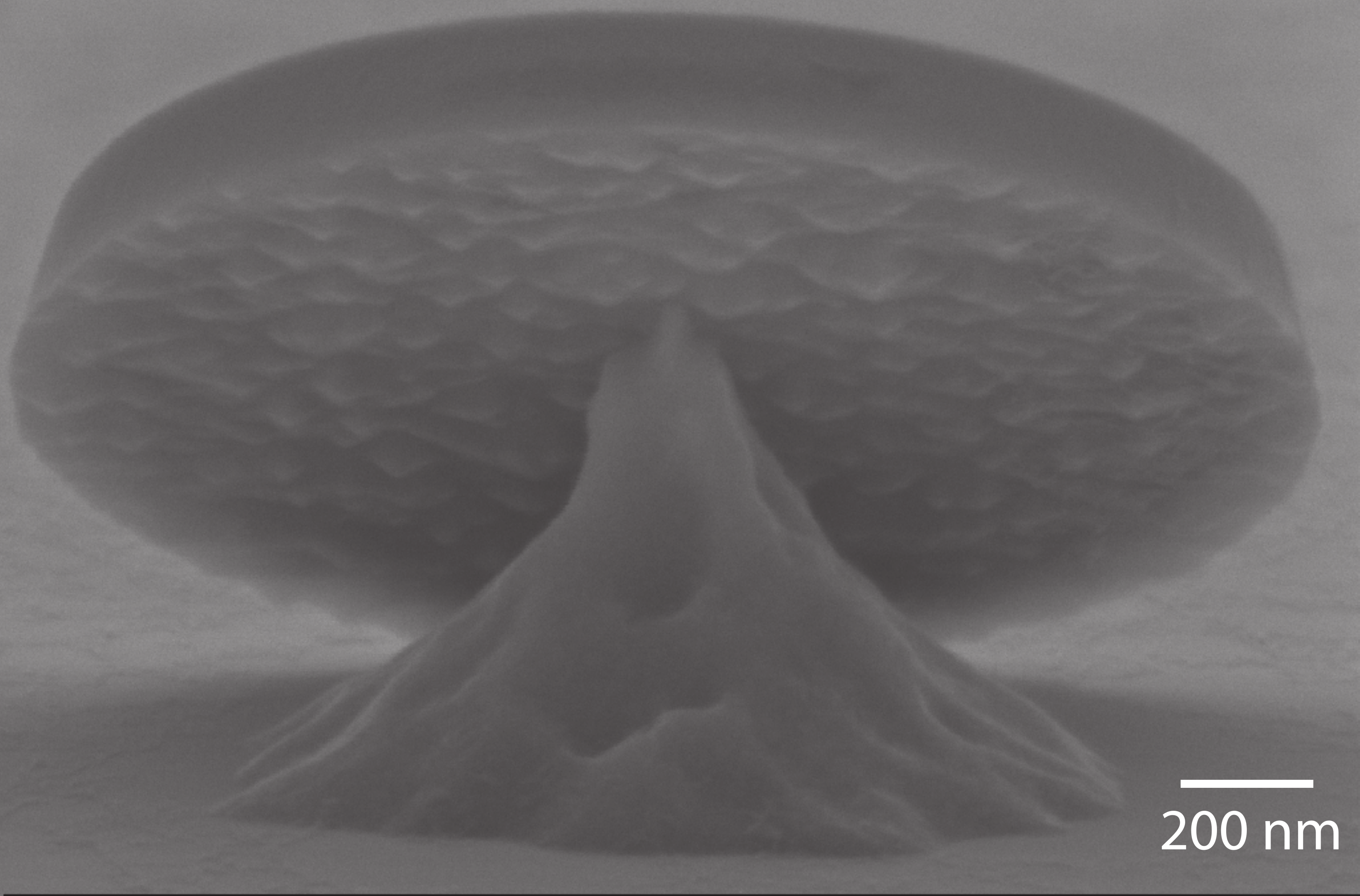}}
 \caption{Scanning electron microscope image of a collapsed microdisk showing roughness on the underside of the GaAs layer.}
\label{fig:FigureSI_SEM}
\end{figure}

The low optical quality factors measured for the GaAs microdisks in this work ($<5{\times}10^4$) relative to those measured in other work using the same fabrication process~\cite{ref:Srinivasan9} can be attributed to the presence of significant underside roughness in the GaAs layer which causes additional scattering. The roughness can be seen clearly in Fig.~\ref{fig:FigureSI_SEM}, which shows the underside of a collapsed microdisk. The roughness is apparently unrelated to the lithography and dry etching processes, since the underside of the GaAs layer is protected during these stpdf.  This suggests that either the wet undercut step, or the initial growth itself, is where the roughness originates.  Given the success of the same wet undercut procedure in fabricating similar GaAs devices by many research groups, our current hypothesis is that the roughness arises in the underlying Al$_{0.6}$Ga$_{0.4}$As layer, possibly due to temperature variations during growth, and acts as a template for the subsequent GaAs layer growth.  This hypothesis is consistent with other reports in the literature~\cite{ref:sweet2010gaas}.

\section*{Phase modulator calibration}

We use a phase modulator to calibrate the optomechanical coupling rate $g_{0}$. The basic idea
is to relate the modulation produced by sending light through the cavity optomechanical system (which
is driven by its contact with the thermal environment) to a direct phase modulation applied with an
electro-optic phase modulator.  Because both undergo the same transduction function, as shown by
Gorodetsky et al.~\cite{ref:Gorodetsky_Kippenberg_OM}, the ratio of the integrated power in their photocurrent
RF spectra will be related to the ratio of $g_{0}$ and the phase modulator's modulation index.  We go through
this derivation below.

The root-mean-square (rms) thermal displacement amplitude ($\alpha
_{thermal}$) is related to the temperature ($T$) using the equipartition theorem:%

\[
\frac{1}{2}m_{eff}\Omega_{m}^{2}\alpha_{thermal}^{2}=\frac{1}{2}k_{B}T
\]

\noindent where $m_{eff}$ is the motional mass of the oscillator and $\Omega_{m}$ is the
mechanical mode frequency.  We define a thermal modulation index~\cite{ref:Eichenfield_thesis}:%

\[
\beta_{thermal}=\frac{\alpha_{thermal}g_{om}}{\Omega_{m}}%
\]

\noindent where the optomechanical coupling parameter ($g_{om}$) is defined by:%

\[
\omega(\alpha)=\omega_{0}+g_{om}\alpha
\]

\noindent with $\omega_{0}$ being the unperturbed cavity frequency.  This leads to:%

\[
\beta_{thermal}^{2}=\frac{k_{B}T}{m_{eff}\Omega_{m}^{2}}\frac{g_{om}^{2}%
}{\Omega_{m}^{2}}%
\]

The vacuum optomechanical coupling rate $g_{0}$~\cite{ref:Gorodetsky_Kippenberg_OM} is:

\[
g_{0}=g_{om}x_{zpf}%
\]

\noindent where the (rms) amplitude of the zero point fluctuation is:%

\[
x_{zpf}=\sqrt{\frac{\hbar}{2m_{eff}\Omega_{m}}}%
\]

\noindent which can be directly calculated as $\sqrt{\bra{0} x^2 \ket{0}}$ from the ground state wavefunction $\ket{0}$ of
the simple harmonic resonator.

We then relate $g_{0}$ to $\beta_{thermal}$:%

\[
\beta_{thermal}^{2}=\frac{2k_{B}T}{\hbar\Omega_{m}^{3}}g_{0}^{2}%
\]

\noindent The modulation index ($\beta_{pm}$) of the phase modulator is defined as:%

\[
\beta_{pm}=\frac{\pi V_{sig}}{V_{\pi}}%
\]

\noindent where $v_{sig}$ is the signal amplitude and $v_{\pi}$ is the modulator half
wave voltage.

By comparing the (integrated) powers in the cavity mechanical mode signal and
the phase modulator signal obtained from the electronic spectrum analyzer, we get:%

\[
g_{0}^{2}=\frac{\hbar\Omega_{m}}{2k_{B}T}\Omega_{m}^{2}\beta_{pm}^{2}%
\frac{S_{cav}(\Omega_{m})}{S_{pm}(\Omega_{pm})}%
\]

\section*{Phase modulator V$_{\pi}$ measurement}

\begin{figure}[h]
\centerline{\includegraphics[width=0.5\linewidth]{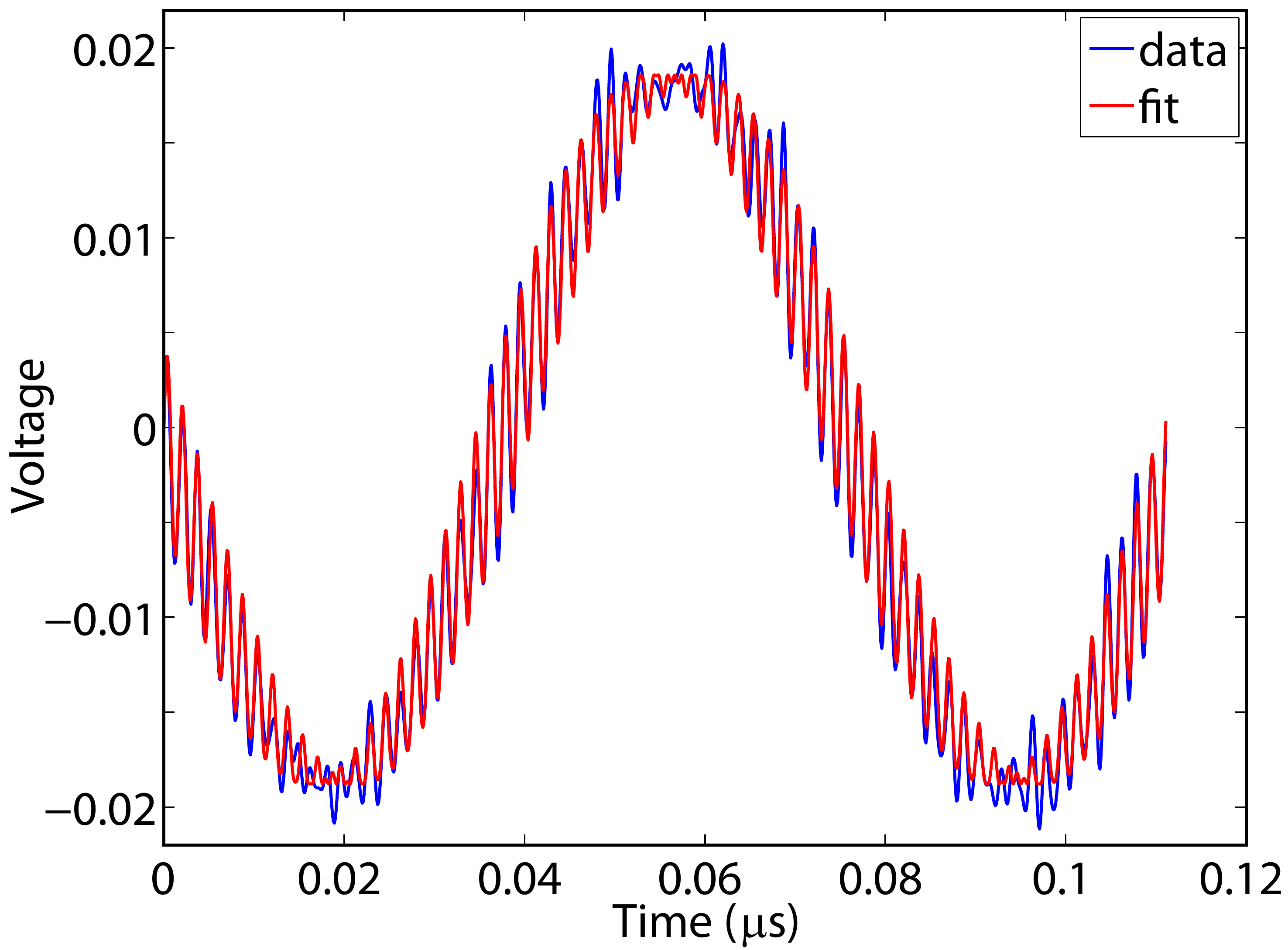}}
 \caption{Calibration of the phase modulator V$_{\pi}$.  Plot shows the measured oscilloscope trace (blue) and the fit (red) using eqn.~\ref{eq:Vpi_eqn}, for a 600 MHz RF signal applied to the phase modulator.}
\label{fig:FigureSI_2}
\end{figure}

Calibration of the phase modulator $V_{\pi}$ is an important step in determining the optomechanical coupling rate $g_{0}$ using the procedure outlined above.  Following the approach and notation described in~\cite{ref:weis_thesis}, we use an unbalanced, fiber Mach-Zehnder interferometer with the phase modulator and input polarization controller placed in one of the arms. The 1550~nm laser is scanned across the interferometer and the transmitted signal is monitored on an oscilloscope that is trigerred with a period corresponding to the laser scan rate. Applying an RF drive to the phase modulator induces modulation sidebands which show up as an amplitude modulation (at the modulation frequency) on the transmitted signal after it passes through the interferometer. The transmitted signal is given as:
\begin{equation}
V(t)=V_{0}+\Delta V_{2}\sin(\frac{2\pi}{\Delta t_{1}}+\Delta\phi_{\max}%
\sin(\frac{2\pi}{\Delta t_{2}}t+\phi_{1}))
\label{eq:Vpi_eqn}
\end{equation}

\noindent where $V_{0}$ is a background offset, $\Delta V_{2}$ is the amplitude of the sinusoidal modulation, $\Delta\phi_{\max}={\frac{{\pi}V_{sig}}{V_{\pi}}}$ is the maximum phase modulation which corresponds to the peak voltage of the applied RF signal, $\Delta t_{1}$ is the inverse of the laser scan speed, and $\Delta t_{2}$ is the inverse of the modulation frequency. A nonlinear least squares fit of the transmitted voltage signal from the oscilloscope to eqn.~\ref{eq:Vpi_eqn} is used to extract $V_{\pi}$. A representative spectrum and fit for a 600 MHz RF signal are shown in Fig. ~\ref{fig:FigureSI_2}, where the $V_{\pi}$ = 3.94~V$~\pm~$0.2~V. This value is in good agreement with that specified by the vendor datasheet (4~V).

\section*{Fiber taper influence on the optomechanical coupling rate}

\begin{figure}[h]
\centerline{\includegraphics[width=0.5\linewidth]{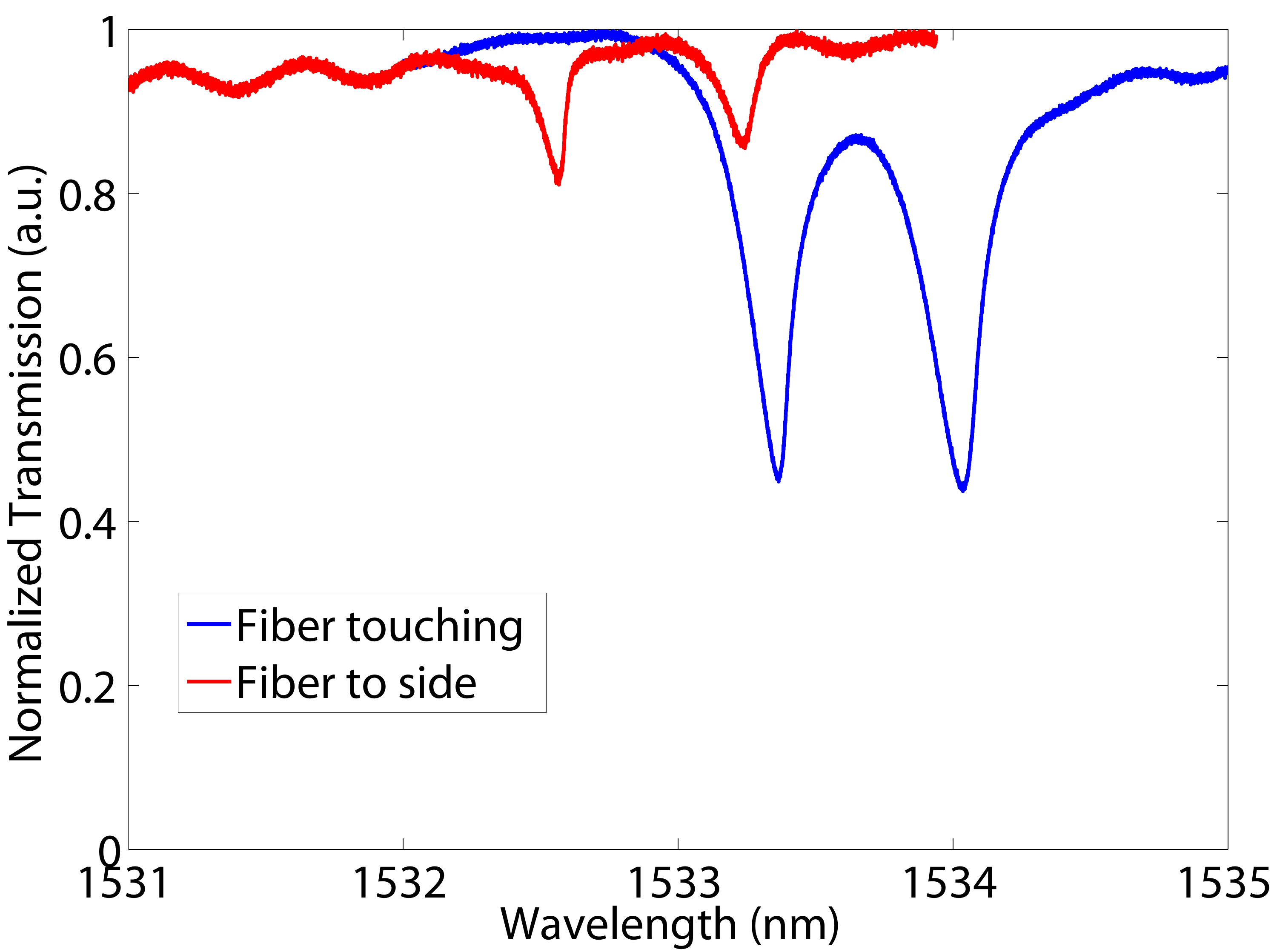}}
 \caption{Transmission spectrum of a $R=1.25$~$\mu$m microdisk probed using the fiber taper waveguide touching the disk (blue) and off to the side(red).}
\label{fig:FigureSI_opt_mode_taper}
\end{figure}

\begin{figure}[h]
\centerline{\includegraphics[width=0.5\linewidth]{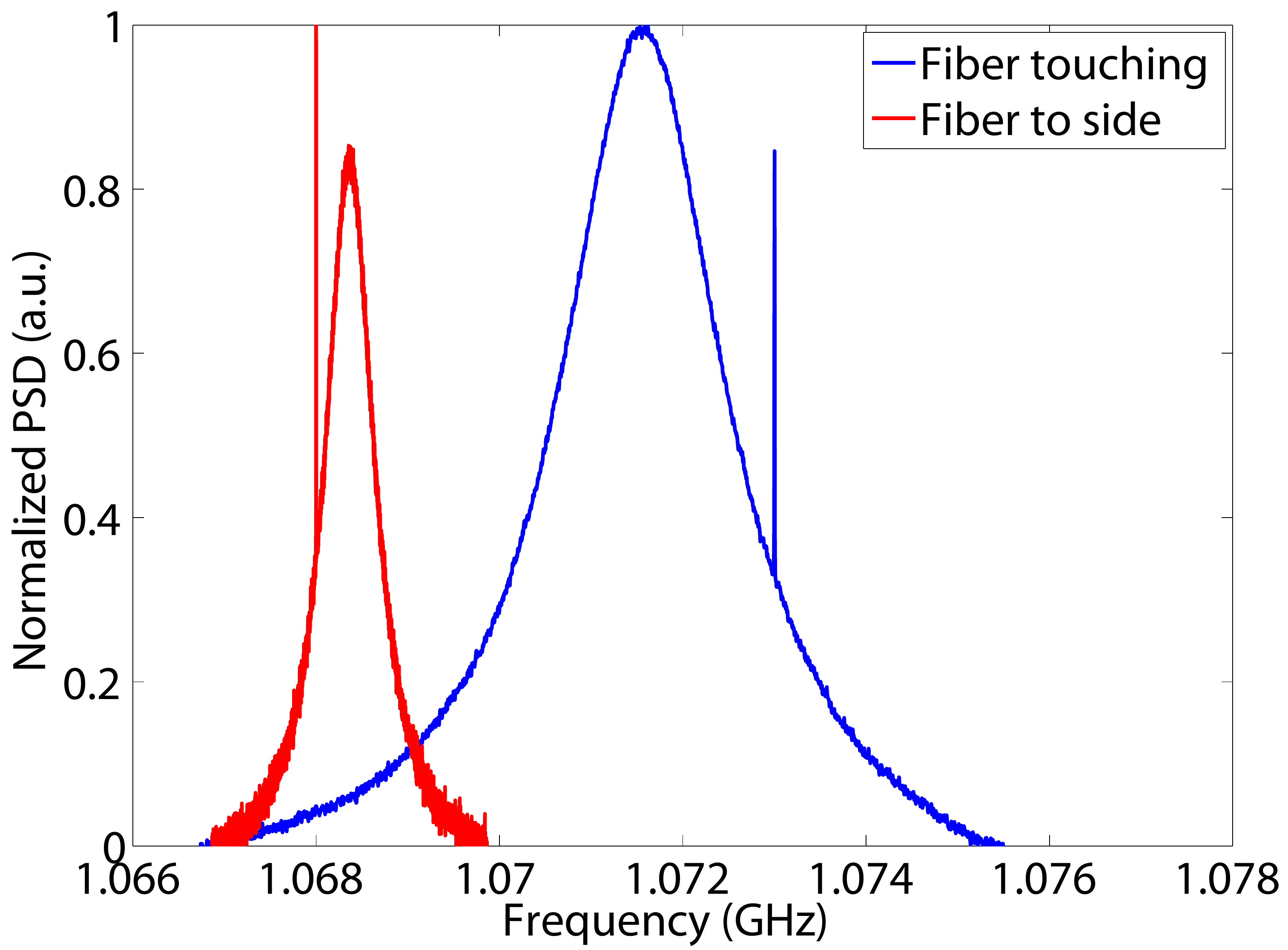}}
 \caption{Normalized mechanical mode power spectral density of a $R=1.25$~$\mu$m microdisk probed using the fiber taper touching the disk (blue) and off to the side (red).}
\label{fig:FigureSI_mech_mode_taper}
\end{figure}

The optomechanical coupling rate $g_{0}$ in the microdisk is influenced by the presence of the fiber taper waveguide in the near field. Given the high refractive index of GaAs and the accompanying tight mode confinement, the fiber taper waveguide has to be brought in close proximity to the disk to ensure sufficient coupling depth so that one can observe the mechanical modes. This problem is exacerbated in our case by the underside roughness which lowers the intrinsic optical $Q$ of the cavities significantly.  If the fiber taper comes in contact with the disk, it perturbs both the optical mode (by modifying the field distribution) and the mechanical mode (by clamping the motion where the fiber touches the disk). This can be seen by a shift in the resonance wavelength of the disk in both the optical (Fig.~\ref{fig:FigureSI_opt_mode_taper}) and mechanical mode spectra (Fig.~\ref{fig:FigureSI_mech_mode_taper}) as well as an increase in the corresponding damping rates.  We find that the extracted coupling strength for devices in which the fiber taper has come into contact with the microdisk can be significantly different ($\approx25~\%$ in some cases) than those in which the fiber is held to the side of the disk.  For the measurements shown in Fig.~2 in the main text, we have attempted to keep the fiber taper waveguide coupling position consistent for all of the measured microdisks, with the coupling depth (extinction ratio $<$ 75~\%) of the optical mode and the mechanical Q ($>$ 1000) of the radial breathing mode mode serving as consistency checks for disks of the same nominal diameter.

Finally, we note that while our work has assumed that the inferred optomechanical coupling rate $g_{0}$ is purely due to dispersive coupling (consisting of both moving dielectric boundary and photoelastic components), recent work has highlighted the potential for other coupling mechanisms, such as dissipative coupling in which the intrinsic and extrinsic quality factors of the optical cavity depend on the motion of the mechanical resonator~\cite{ref:Wu_Barclay_dissipative_dispersive_OM}.  These authors have found that the fiber taper waveguide can influence the magnitude of both the dissipative and dispersive optomechanical coupling rates~\cite{ref:Hryciw_Barclay_arXiv}.

\section*{Optomechanical coupling for the TE$_{2,m}$ microdisk mode}

\begin{figure}[h]
\centerline{\includegraphics[width=0.5\linewidth]{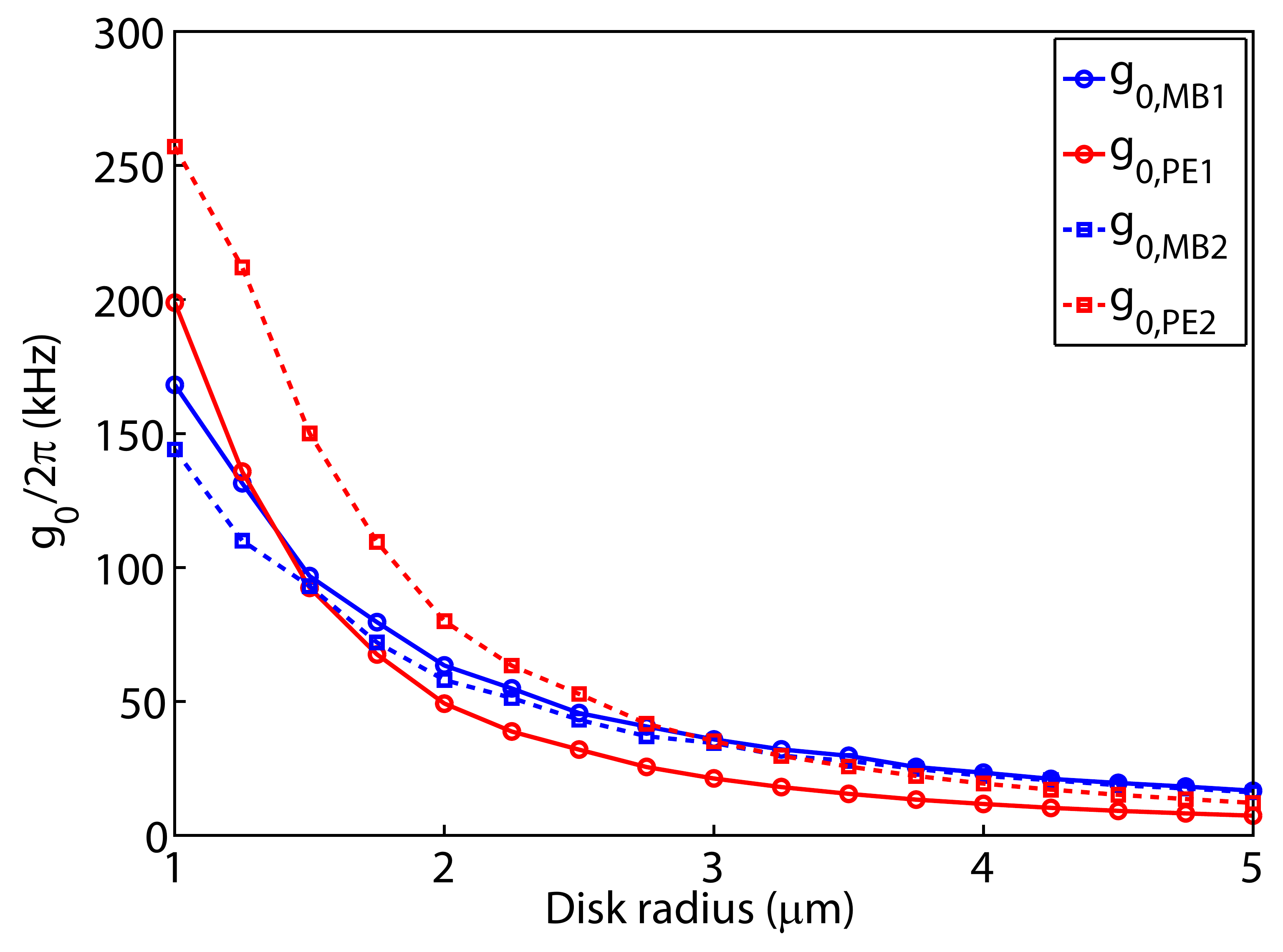}}
 \caption{Calculated moving boundary (blue) and photoelastic effect (red) contributions to the optomechanical coupling rate between the TE$_{1,m}$ (solid) and TE$_{2,m}$ (dashed) whispering gallery optical modes and the first order radial breathing mechanical mode.}
\label{fig:FigureSI_te2m}
\end{figure}

We have also considered the moving boundary and photoelastic components to the optomechanical coupling rate $g_{0}$ between the TE$_{2,m}$ whispering gallery mode and radial breathing mechanical mode in GaAs microdisks.  Simulation results are shown in Fig.~\ref{fig:FigureSI_te2m}. For comparison, we also plot the corresponding values for the TE$_{1,m}$ mode. As can be seen, the moving boundary contribution to $g_{0}$ remains comparable for the two modes but the photoelastic contribution for TE$_{2,m}$ is greater than that for the TE$_{1,m}$ mode. Since the photoelastic contribution roughly scales as $pS|E|^{2}$, the higher coupling for the TE$_{2,m}$ can be attributed to greater overlap between the electric field and the displacement in the interior of the disk.  In practice, we did not look for TE$_{2,m}$ due to their relatively low radiation-limited optical quality factors for the smallest diameter disks.

\section*{Silicon nanobeam optomechanical crystal design}

\begin{figure}
\begin{center}
\includegraphics[width=0.5\linewidth]{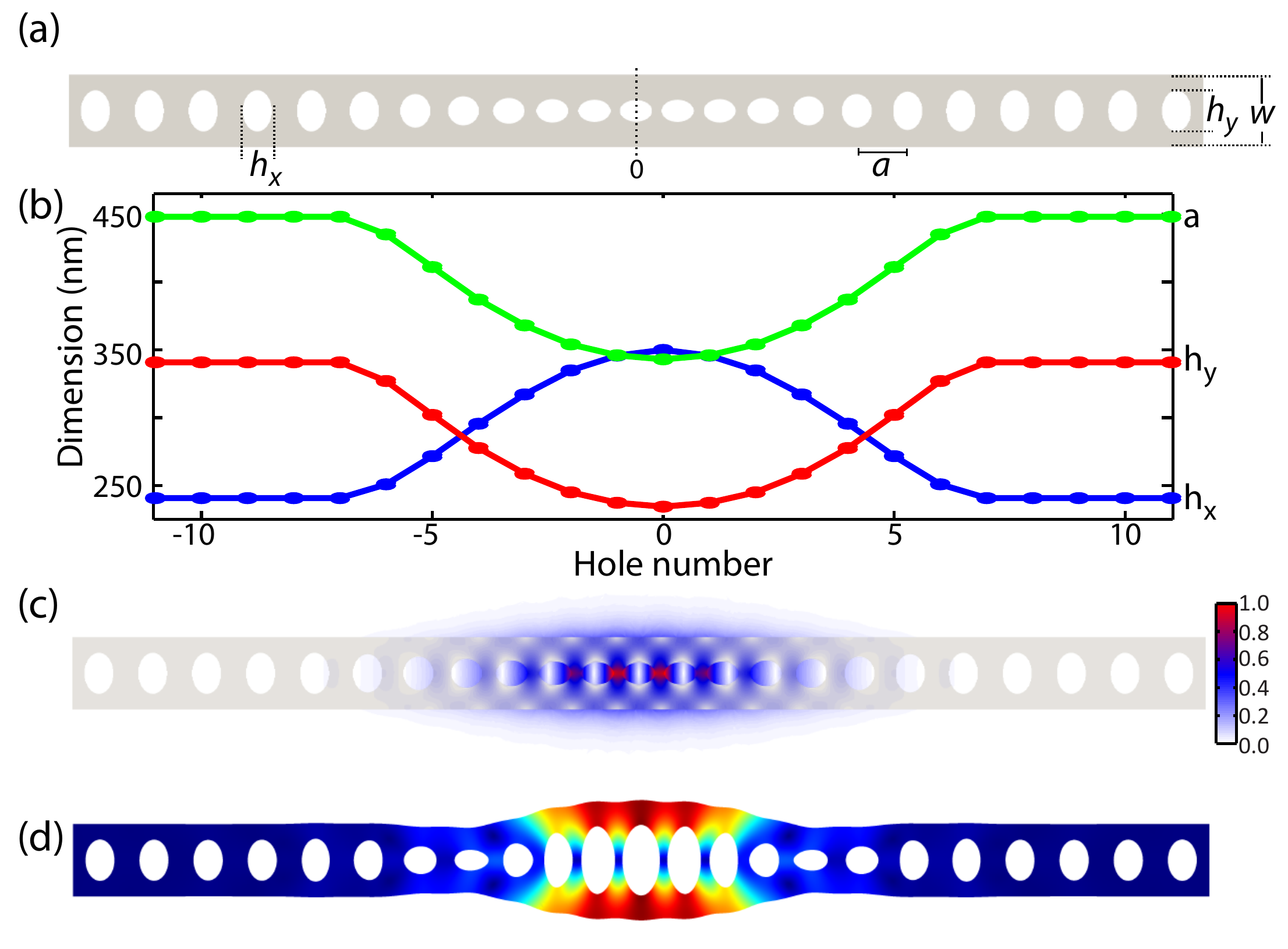}
\caption{Si nanobeam optomechanical crystal (a) design, (b) variation in design parameters as a function of hole number, (c) normalized electric field amplitude, and (d) normalized mechanical displacement field.}\label{fig:FigureSI_Si_NB}
\end{center}
\end{figure}

The schematic, beam parameters, normalized electric field amplitude of the optical mode, and normalized mechanical mode displacement of the Si nanobeam design discussed in Fig. 4 in the main text is shown in Fig.~\ref{fig:FigureSI_Si_NB}. The Young's modulus, density and refractive index values used were 170 GPa, 2329 kg/m$^{3}$, and 3.48 respectively. The nanobeam design is based on Chan et al.~\cite{ref:chan_optimized_OMC}.

\section*{Rotated photoelastic tensor}

The rotated photoelastic tensor can be constructed following ~\cite{ref:chan2012_phd} ($r$
subscript indicates component in rotated frame, $\theta$ is the in-plane
rotation angle from [100] ):%

\[
p_{11r}=p_{12r}=\frac{1}{4}(p_{11}(3+\cos(4\theta))+(p_{12}+2p_{44}%
)(1-\cos(4\theta)))
\]

\[
p_{12r}=p_{21r}=\frac{1}{4}(p_{12}(3+\cos(4\theta))+(p_{11}-2p_{44}%
)(1-\cos(4\theta)))
\]

\[
p_{33r}=p_{11},p_{13r}=p_{23r}=p_{31r}=p_{32r}=p_{12}%
\]

\[
p_{44r}=p_{55r}=p_{44}
\]
\[
p_{66r}=\frac{1}{4}(2p_{44}(1+\cos(4\theta))+(p_{11}-p_{12})(1-\cos(4\theta)))
\]

\[
p_{16r}=p_{61r}=\frac{1}{4}\sin(4\theta)(2p_{44}+p_{12}-p_{11})
\]

\[
p_{26r}=p_{62r}=\frac{1}{4}\sin(4\theta)(p_{11}-p_{12}-2p_{44})
\]

The $p\cdot S$ term can be constructed as:%

\[
\left[
\begin{array}
[c]{cccccc}%
p_{11r} & p_{12r} & p_{13r} & 0 & 0 & p_{16r}\\
p_{21r} & p_{22r} & p_{23r} & 0 & 0 & p_{26r}\\
p_{31r} & p_{32r} & p_{33r} & 0 & 0 & 0\\
0 & 0 & 0 & p_{44r} & 0 & 0\\
0 & 0 & 0 & 0 & p_{55r} & 0\\
p_{61r} & p_{62r} & 0 & 0 & 0 & p_{66r}%
\end{array}
\right]  \left[
\begin{array}
[c]{c}%
S_{1}=S_{xx}\\
S_{2}=S_{yy}\\
S_{3}=S_{zz}\\
S_{4}=2S_{yz}\\
S_{5}=2S_{xz}\\
S_{6}=2S_{xy}%
\end{array}
\right]  =\left[
\begin{array}
[c]{c}%
pS_{1}\\
pS_{2}\\
pS_{3}\\
pS_{4}\\
pS_{5}\\
pS_{6}%
\end{array}
\right]
\]

and the perturbation can be calculated as:%

\[
\frac{d\omega}{d\alpha}=\frac{\omega_{0}\epsilon_{0}n^{4}}{2}\frac{\int%
_{GaAs}dx\left[
\begin{array}
[c]{ccc}%
E_{x}^{\ast} & E_{y}^{\ast} & E_{z}^{\ast}%
\end{array}
\right]  \left[
\begin{array}
[c]{ccc}%
pS_{1} & pS_{6} & pS_{5}\\
pS_{6} & pS_{2} & pS_{4}\\
pS_{5} & pS_{4} & pS_{3}%
\end{array}
\right]  \left[
\begin{array}
[c]{c}%
E_{x}^{{}}\\
E_{y}^{{}}\\
E_{z}^{{}}%
\end{array}
\right]  }{\int dx\epsilon|E|^{2}}%
\]

\section*{Breathing mode simulation with aniostropic elastic constants}

\begin{figure}[h]
\centerline{\includegraphics[width=0.5\linewidth]{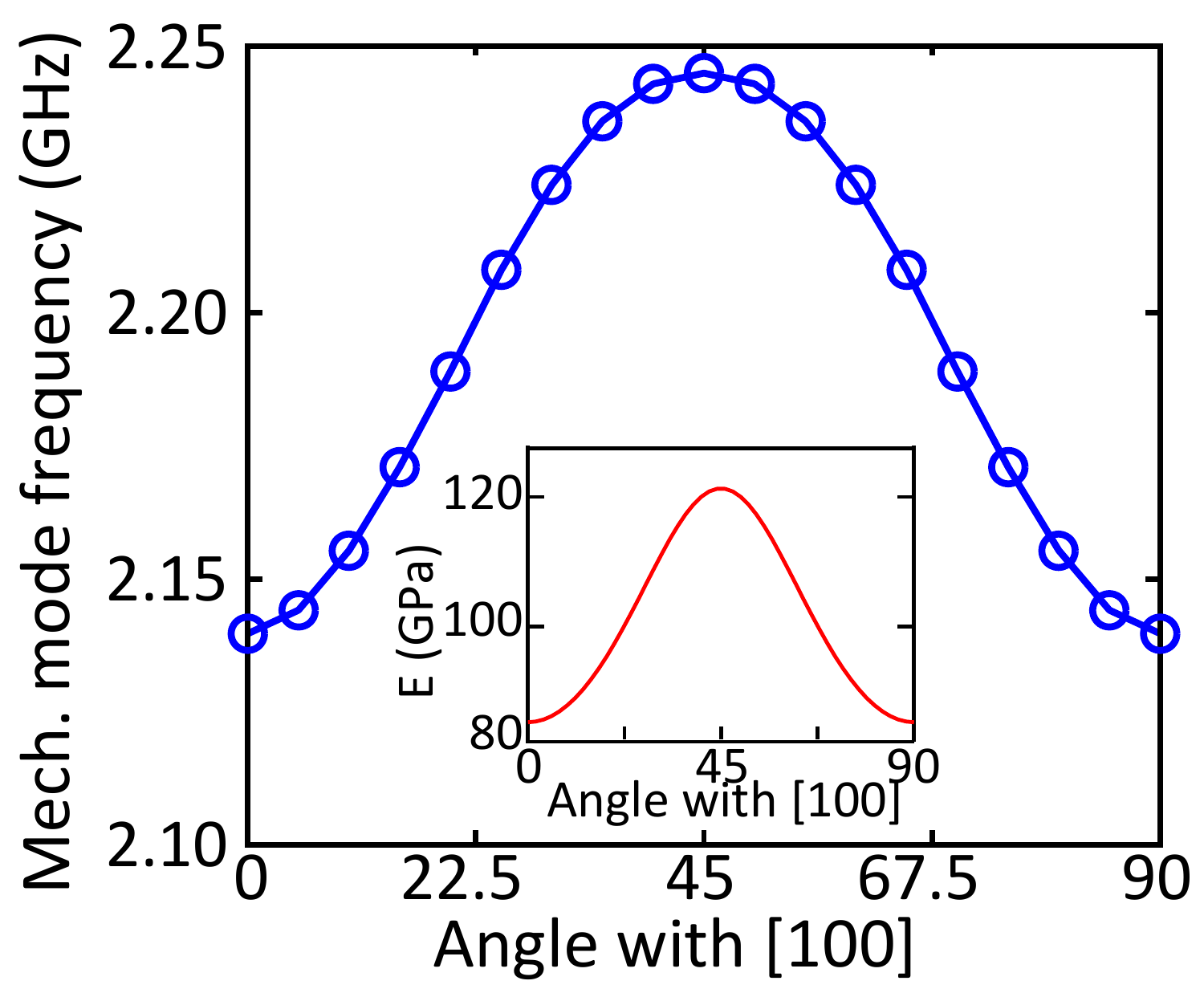}}
 \caption{Dependence of the GaAs nanobeam breathing mode frequency on the in-plane orientation (defined with respect to the long axis of the nanobeam). The inset shows the corresponding orientation dependence of the Young's modulus of GaAs.}
\label{fig:FigureSI_rbm_ang}
\end{figure}

The elastic anisotropy of GaAs ($s_{11}=1.173$, $s_{12}=-0.366$ and $s_{44}=1.684$ (in units of $10^{-13}$ Pa)~\cite{ref:brantley_E_fn_angle,ref:Hopcroft_Youngs_Modulus_Si} leads to an orientation-dependent Young's modulus (inset of Fig.~\ref{fig:FigureSI_rbm_ang}), Poisson's ratio, and shear modulus, and a corresponding orientation dependence of the nanobeam mechanical breathing mode frequency. Figure~\ref{fig:FigureSI_rbm_ang} shows this dependence, where a frequency shift of $\approx$~100~MHz is expected between devices oriented along [100] and [110].  This corresponds reasonably with the measured frequency shift of $\approx$~85~MHz.

\begin{figure}[h]
\centerline{\includegraphics[width=0.5\linewidth]{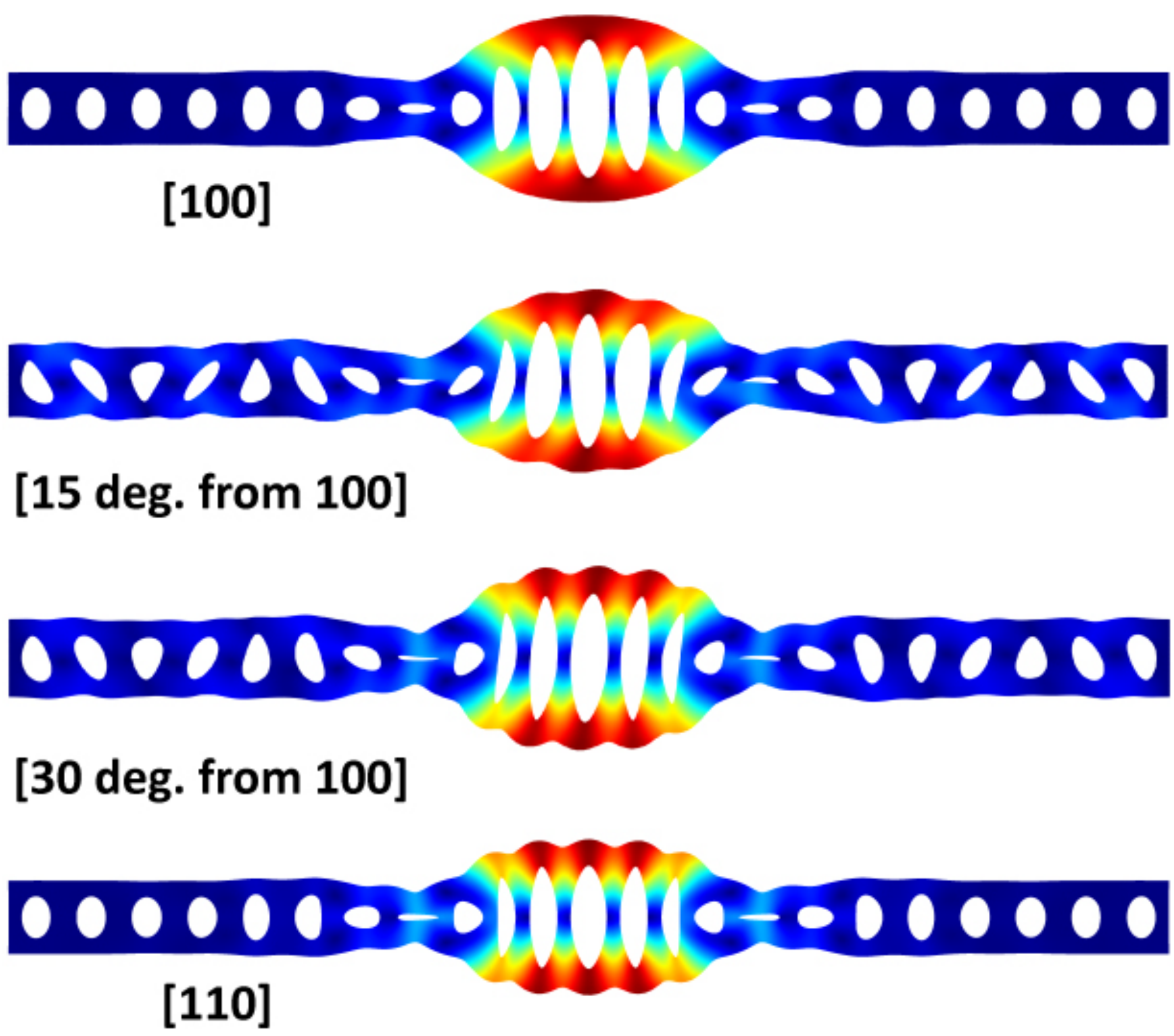}}
 \caption{GaAs nanobeam optomechanical crystal breathing mode profiles for nanobeam long axis oriented at 0~$^{\circ}$, 15~$^{\circ}$, 30~$^{\circ}$, and 45~$^{\circ}$ with respect to [100], respectively.}
\label{fig:FigureSI_rbm_ang_mode_plots}
\end{figure}

For calculating the breathing mode frequency as a function of in-plane angle, GaAs was represented as an anisotropic material using an orthotropic elasticity matrix with parameters $E_{x}$ = 121.2 GPa, $E_{y}$ = 121.2 GPa, $E_{z}$ = 85.9 GPa, $\nu_{xy}$ = 0.0209, $\nu_{yz}$ = 0.4434, $\nu_{xz}$ = 0.312, $G_{xy}$ = 32.5 GPa, $G_{yz}$ = 59.4 GPa, and $G_{xz}$ = 59.4 GPa. The values used here correspond to the x-axis along [110]. The beam is physically rotated about the z-axis to calculate the breathing mode frequency as a function of in-plane angle. Figure~\ref{fig:FigureSI_rbm_ang_mode_plots} shows the displacement profile for the breathing mode for 0~$^{\circ}$, 15~$^{\circ}$, 30~$^{\circ}$, and 45~$^{\circ}$ from the [100] axis. For 15~$^{\circ}$ and 30~$^{\circ}$, the displacements are no longer symmetric (as can be seen from the distorted hole shapes) and the mechanical mode confinement is reduced. As a general rule of thumb, designs along [110] or [100] are expected to give the highest mechanical Q.

\section*{Dependence of the moving boundary contribution on in-plane orientation}

The rotation dependent $g_{0,PE}$ shown in Fig. 4 in the main text was calculated assuming GaAs was an isotropic material ($E$ = 85.9 GPa, ${\nu}$=0.31, ${\rho}=5317$) and rotating the photoelastic tensor. For an isotropic material, $g_{0,MB}$ is independent of in-plane orientation as can be seen from the form of eqn. 1 in the main text.

For completeness, we calculated $g_{0,MB}$ and $g_{0,PE}$ for nanobeams simulated with GaAs as an orthotropic elastic material (with mode shapes shown in Fig.~\ref{fig:FigureSI_rbm_ang_mode_plots}). For calculating the overlap integrals along [100], we rotated the nanobeam by 45~$^{\circ}$ and solved for the optical mode. As expected from the different breathing mode profiles, $g_{0,MB}$ is different along [100] ($g_{0,MB}/2\pi=-73$~kHz) and [110] ($g_{0,MB}/2\pi=-15$~kHz). The total $g_0$ along [100] ($g_{0}/2\pi=$~850~kHz) and [110] ($g_{0}/2\pi=$~1.06~MHz) are within ten percent of the values reported in Fig. 4(a) in the main text, on account of being dominated by the photoelastic effect.


\end{document}